\documentclass[5p,twocolumn]{elsarticle}
\usepackage{graphicx,latexsym}
\usepackage{dcolumn}
\usepackage{amssymb,amsmath,bm}
\usepackage{subfigure}
\usepackage{braket}
\usepackage{multicol}
\usepackage{siunitx}
\usepackage{array}
\usepackage{float}
\usepackage[nopar]{lipsum}
\usepackage{caption}
\usepackage{multirow}
\usepackage{bigstrut}
\usepackage[super]{nth}

\newcommand{\angstrom}{\text{\normalfont\AA}}
\usepackage{hyperref}
\hypersetup{
  pdfnewwindow=true,       
  colorlinks=true,         
   linkcolor=blue,          
   citecolor=blue,          
   filecolor=magenta,       
   urlcolor=black           
}

\usepackage[normalem]{ulem}

\def\sec#1{Sec.\ \ref{#1}}

\def\fig#1{Fig.\ \ref{#1}}
\def\tab#1{Tab.\ \ref{#1}}

\journal{}

\begin{document}

\begin{frontmatter}


\title{Spin-polarised DFT modeling of electronic, magnetic, thermal and optical properties of 
       silicene doped with transition metals}

\author[a1,a2]{Nzar Rauf Abdullah}
\ead{nzar.r.abdullah@gmail.com}
\address[a1]{Division of Computational Nanoscience, Physics Department, College of Science, 
             University of Sulaimani, Sulaimani 46001, Kurdistan Region, Iraq}
\address[a2]{Computer Engineering Department, College of Engineering, Komar University of Science and Technology, Sulaimani 46001, Kurdistan Region, Iraq}

\author[a3]{Mohammad T. Kareem}
\address[a3]{Chemistry Department, College of Science, University of Sulaimani, 
	         Sulaimani 46001, Kurdistan Region, Iraq}

\author[a1]{Hunar Omar Rashid}              

\author[a4]{Andrei Manolescu}
\address[a4]{Reykjavik University, School of Science and Engineering,
	Menntavegur 1, IS-101 Reykjavik, Iceland}

\author[a5]{Vidar Gudmundsson}   
\address[a5]{Science Institute, University of Iceland,
	Dunhaga 3, IS-107 Reykjavik, Iceland}

%

\begin{abstract}
	
The geometric, electronic, magnetic, thermal, and optical properties of transition metal (TM) doped silicene are systematically explored using spin-dependent density functional computation. 
We find that the TM atoms decrease the buckling degree of the silicene structure caused by  
the interaction between the dopant TM atoms and the Si atoms in the silicene layer plane
which is quite strong.
In some TM-silicenes, parallel bands and the corresponding van Hove singularities are observed 
in the electronic band structure without and with spin-polarization. 
These parallel bands are the origin of most of the transitions in the visible and the UV regions.
A high Seebeck coefficient is found in some TM-silicene without spin-polarization. 
In the presence of emergent spin-polarization, a reduction or a magnification of the Seebeck coefficient is seen due to a spin-dependent phase transition. 
We find that the preferred state is a ferromagnetic state with a very 
high Curie temperature.
We observe a strong interaction and large orbital hybridization between 
the TM atoms and the silicene. As a result, a high magnetic moment emerges in TM-silicene. 
Our results are potentially beneficial for thermospin, and optoelectronic nanodevices.

\end{abstract}

\begin{keyword}
Magnetization \sep Thermal transport \sep Silicene \sep DFT \sep Electronic structure \sep  Optical properties
\end{keyword}

\end{frontmatter}

\section{Introduction}

A two-dimensional allotrope of silicon is silicene, which was first reported in 1994 \cite{PhysRevB.50.14916}. Silicene has a periodically buckled topology leading to different properties compared to some other 2D materials~\cite{Tao2015,Jia2016}, and it is composed of silicon (Si) atoms with 
some benefits compared to graphene \cite{ZHAO201624, RASHID2019102625} as it is compatible with the present silicon-based technology.
Therefore, silicene has been extensively investigated in electronics \cite{LeLay2015}, optical \cite{Do2019}, thermoelectric  \cite{Fu2015}, and magnetic \cite{Tokmachev2018} devices. 
Silicenes with Transition metals (TM) have also several applications in chemistry such as hydrogenation evolution reaction \cite{C9RA04602J}, designing oxygen reduction reaction electro-catalysts \cite{C9CP04284A}, and silicene superlattice for Na-ion \cite{Zhu_2016} Li-O$_2$ \cite{zhang2018lithiation} batteries.

Despite its unique properties, silicene is not a very good material for some applications such as thermoelectric devices due to the zero bandgap. However, the zero-gap disadvantage can be overcome~\cite{Sadeghi2015}. Transition metals (TM) doped silicene (TM-silicene) are candidates in which the bandgap can be tuned \cite{PhysRevB.88.165422}. It has been shown that in TM the semi-metal characteristics of silicene are changed to be semiconducting or metallic depending on the type of the TM dopant atoms.
Among the ten types of TM-silicene investigated, Ti-, Ni-, and Zn-doped silicene have shown semiconducting properties, whereas Co- and Cu- doped silicene present a half-metallic material  \cite{doi:10.1002/cphc.201402613}.

In addition to the advantage found in the aforementioned studies, another point speaking for 
TM-silicene is that it is a strong candidate for the quantum spin Hall effect. 
This is again attributed to the enhanced bandgap of TM-silicene 
\cite{Zhang2013, PhysRevLett.107.076802}. 
Therefore, the investigations of the magnetic properties of TM-silicene
are of potential importance in diverse fields such as quantum
electronics, spintronics, and optoelectronics \cite{Zhang2013, Wolf1488}.
It has been shown that the magnetic behavior of silicene can also be tuned
by different TM dopant atoms \cite{doi:10.1063/1.4921699}. The magnetic modifications mainly come
from the $3$d orbitals of the TM dopant atoms along with a partial contribution from 
the adjacent Si atoms.

The optical properties of silicene is another important aspect of research because it
also has many applications in the optoelectronic industries \cite{doi:10.1063/1.4921699, ABDULLAH2020113996}.
It has been reported that optical response of silicene is in the IR and visible regions \cite{JOHN2017307}. The absorption spectra of P and Al-doped silicenes show
higher absorption compared to pristine silicene, and no important changes in 
the electrical conductivity are found with the doping concentration for in-plane 
light polarization \cite{C4RA07976K}. 
The optical properties of TM-doped silicene have shown that 
the intensity of the absorption peaks decreases for out of the 
plane light polarization \cite{Hasanzadeh_Tazeh_Gheshlagh_2020}.

The thermoelectric characteristics of pristine silicene can also be improved by doping. For example, the presence of defects may decrease the phononic thermal conductance. This enhances the thermoelectric efficiency or the figure of merit $ZT$ \cite{ZHAO20161}. A substitutional B/N doping \cite{abdullah2020properties} and di-hydrogenation \cite{C4CP01039F} of silicene have been regarded as an effective way to enhance the thermoelectric efficiency. All the methods used to increase the thermoelectric efficiency focus on
tuning the bandgap of the silicene structure.

In this work, we consider the effects of TM doping on the physical properties of silicene.
We first study the electronic, thermal and optical characteristics of TM-silicene.
We then use a TM dopant as a prototype magnetic impurity to show that it is possible 
to achieve magnetic properties such as ferro- or antiferromagnetism. 
We will show how magnetically active phases in TM-silicene can enhance the thermoelectric properties
such as the Seebeck coefficient.
It will be shown that TM-doped silicene monolayers could be prominent candidates 
for spintronic devices.

In \sec{Sec:Model} the structure of TM-silicene is briefly over-viewed. In \sec{Sec:Results} the main achieved results are analyzed. In \sec{Sec:Conclusion} the conclusions of the modeling results are presented.

\section{Computational Tools}\label{Sec:Model}

All the calculations of the pure and the TM-silicene properties are performed with 
the Quantum espresso (QE) simulation package \cite{Giannozzi_2009, giannozzi2017advanced}. 
For visualization of the samples, the crystalline and molecular structure the visualization program (XCrySDen) is used \cite{KOKALJ1999176}. 
In QE, the general gradient approximation (GGA) with the Perdew-Burke-Ernzerhof (PBE) potential is employed with a cutoff energy of $1088.45$~eV. In our samples consisting of a $2\times2$ supercell, 
the structures are considered fully relaxed when the Hellmann-Feynman forces are less than $1.2\times10^{-4}$~eV/$\angstrom$, and the total energy changes less than $1.9\times 10^{-4}$~eV. 
The Brillouin zone (BZ) is sampled by a Monkhorst-Pack $k$-mesh of $15\times15\times1$. 
The same $k$-mesh points are used for the SCF calculations. In the density of state (Dos) calculations, a $100\times100\times1$ grids are used.
Furthermore, the Boltzmann transport properties software package (BoltzTraP) is utilized to study the thermal properties of the structures \cite{madsen2006boltztrap-2}. The BoltzTraP code uses a mesh of band energies and has an interface to the QE package \cite{ABDULLAH2020126578}. The optical characteristics of the systems are obtained by the QE code with a broadening of $0.1$~eV.

\section{Results}\label{Sec:Results}

Our results are divided into two groups: First, the results of the spin-independent model, 
the electronic, the thermal and optical properties for TM-silicene. Second, we show the results 
of the spin-dependent model, the electronic band structure, the density of states of TM-silicene, 
and the thermal properties. In addition the magnetic properties including the magnetic moments 
of TM-silicene are presented.

\subsection{Spin-independent calculations}

\begin{figure}[htb]
 	\centering
 	\includegraphics[width=0.4\textwidth]{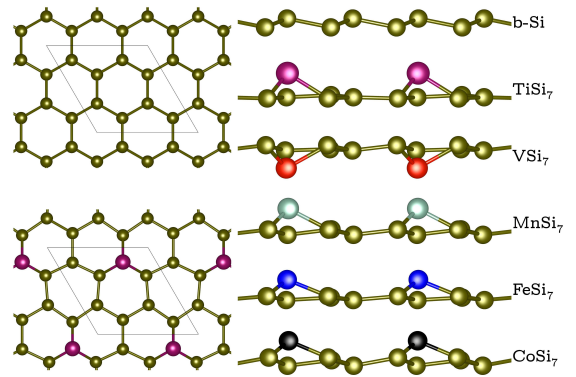}
 	\caption{Left panel: the pure silicene, b-Si, (top panel) and the TM-doped silicene (bottom panel). Right panel: side view of b-Si (golden),  TiSi$_7$ (purple), VSi$_7$ (red), MnSi$_7$ (light blue), FeSi$_7$ (blue), and CoSi$_7$ (black).}
 	\label{fig01}
\end{figure}

In this section, we show the results of the spin-independent model. The structures under investigation are shown in \fig{fig01}. The left panel of \fig{fig01} is the pristine buckled silicene, 
b-Si (top panel), and the TM-doped silicene (bottom panel). In addition, the right panel is the side view of the b-Si and TM-silicene for five selected TM dopants including Ti, V, Mn, Fe, and Co atoms. 
The TM atoms are doped at the para-positions of the $2\times2$ hexagonal structure of silicene \cite{ABDULLAH2020126350,ABDULLAH2020126807, ABDULLAH2020103282}. The TM-silicene is thus identified 
as TiSi$_7$ (purple), VSi$_7$ (red), MnSi$_7$ (light blue), FeSi$_7$ (blue), and CoSi$_7$ (black). 
In b-Si, the Si-Si bond length is found to be $2.27$~$\angstrom$, which agrees well with a previous study \cite{PhysRevB.79.115409}. This larger Si-Si bond length weakens the $\pi\text{-}\pi$ overlaps, resulting in a low-buckled structure. The buckling parameter of b-Si is $0.45$~$\angstrom$ which is 
in a good agreement with previous studies \cite{PhysRevLett.102.236804}.

\begin{table}[h]
	\begin{center}
		\caption{\label{table_1} The buckling parameter of the structures.}
		\begin{tabular}{ |c|c| }    \hline
			Structure     & $\delta (\angstrom)$ \\ \hline
			b-Si          & 0.45     \\ 
			TiSi$_{7}$    & 0.076    \\ 
			VSi$_{7}$     & 0.2      \\ 
			MnSi$_{7}$    & 0.268    \\ 
			FeSi$_{7}$    & 0.174    \\ 
			CoSi$_{7}$    & 0.371    \\ 	
			\hline
		\end{tabular}
	\end{center}
\end{table}

The first observation of \fig{fig01} is that the buckling degree is strongly influenced by the 
TM atom dopant. The buckling parameter ($\delta$) for all structures under investigation are 
presented in \tab{table_1}.

It can be seen that the TM atoms leave the silicene plane in the fully relaxed structures \cite{abdullah2020interlayer}.
Consequently, the buckling degree of TM-silicene is decreased compared to b-Si. 
The outward movement of the TM atoms from the silicene layer and the occupation of almost 
perfectly symmetric 3-fold positions have been reported previously \cite{doi:10.1021/acs.jpclett.7b00115}.
This reveals that in contrast to graphene the
interaction between the TM atoms and the silicene layer is quite strong due to its highly 
reactive buckled hexagonal structure \cite{PhysRevB.87.085423}.
The lowest buckling degree occurring for TiSi$_{7}$ reveals a maximum distortion, 
and the highest buckling degree of CoSi$_7$ among TM-silicene indicates a minimum distortion 
in the silicene structure.
These distortions emerge here because the bonding energy of the Ti atoms, $4.89$~eV, 
to silicene occurs with a significant buckling and lattice a distortion \cite{PhysRevB.87.085423}. 
In addition, the atomic radii of the Ti atoms are much larger than these of the Co atoms which 
may influence the distortion and the buckling degree.

\begin{figure}[htb]
	\centering
	\includegraphics[width=0.4\textwidth]{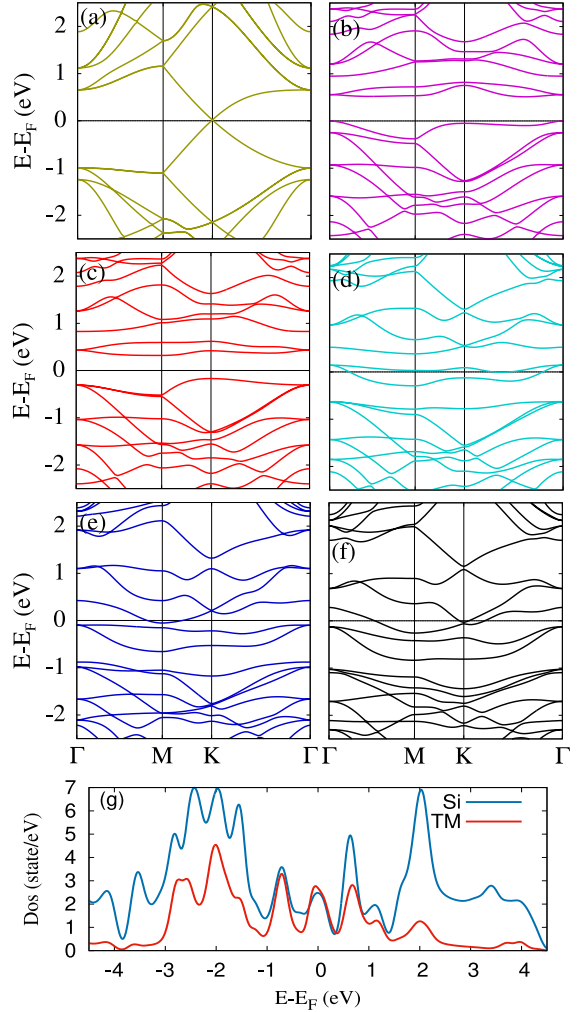}
	\caption{The electronic band structure of b-Si (a), TiSi$_7$ (b), 
		VSi$_7$ (c), MnSi$_7$ (d), FeSi$_7$ (e), and CoSi$_7$ (f). 
		The density of state (Dos) of the Si and TM atoms in TM-silicene, FeSi$_7$, (g).
	    The Fermi energy is set at zero.}
	\label{fig02}
\end{figure}

The DFT calculations of the formation energy indicate that TiSi$_7$ has the lowest, and CoSi$_7$ has the highest formation energy among all the investigated structures. The TiSi$_7$ is thus the most structurally stable system.

The electronic band structures of b-Si (a) and the TM-silicenes (b-f) are presented in \fig{fig02}.
The electronic eigenvalues along high symmetry directions ($\Gamma$ M K $\Gamma$) in the first
Brillouin zone are calculated self-consistently.
Our results for the band structure show that the TiSi$_7$ (b) and VSi$_7$ (c) are semiconducting materials due to the presence of a finite bandgap around Fermi energy, while MnSi$_7$ (c), FeSi$_7$ (e), and CoSi$_7$ (f) are metals because the Fermi energy crosses valence or conduction bands. 
The bandgap of TiSi$_7$ and VSi$_7$ are $0.51$ and $0.59$~eV, respectively.
It should be noticed that the Dirac cone in the TM silicene vanishes. This is caused by the orbitals of the TM atom which make a considerable contribution to the energy levels through the hybridization between the TM atom and the silicene as is seen in \fig{fig02}(g).
It can be clearly seen the density of states of the TM atom around the Fermi energy has 
a high contribution.
The semiconductor properties of some TM-silicenes with Ti dopant atoms have recently been studied \cite{Hasanzadeh_Tazeh_Gheshlagh_2020}.

\begin{figure}[htb]
	\centering
	\includegraphics[width=0.35\textwidth]{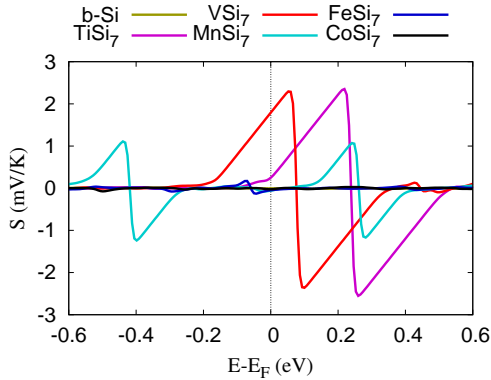}
	\caption{Seebeck coefficient of the b-Si (golden) and the TM-silicene at temperature $T = 100$~K. including TiSi$_7$ (purple), VSi$_7$ (red), MnSi$_7$ (light blue), FeSi$_7$ (blue), and CoSi$_7$ (black). }
	\label{fig03}
\end{figure}

The bandgap tuning directly influences the thermoelectric properties of a system. 
The Seebeck coefficient, $S$, of a pristine b-Si and TM-silicenes are displayed in \fig{fig03} 
at temperature $T = 100$~K.
We focus on this low temperature range from $20$ to $160$~K, where the electrons and phonons 
are decoupled. At this temperature range the electrons deliver the main contribution to the 
thermal behavior \cite{PhysRevB.87.241411,abdullah2019manifestation, abdullah2019thermoelectric}. 
The gapless b-Si exhibits poorer thermoelectric performance, than the gapped TM-silicenes.
The low Seebeck coefficient and the thermoelectric performance of b-Si is caused by 
the cancellation of the electron-hole contributions to the transport quantities.
As we stated before, an effective way to enhance the thermoelectric properties of a system
is to open up a bandgap, and thus lifting this cancellation effect \cite{doi:10.1063/1.5100985}.

We therefore see the maximum Seebeck coefficient for TiSi$_7$ (purple) and VSi$_7$ (red) as they have a bandgap around the Fermi energy behaving as semiconductor materials.

\begin{figure}[htb]
	\centering
	\includegraphics[width=0.40\textwidth]{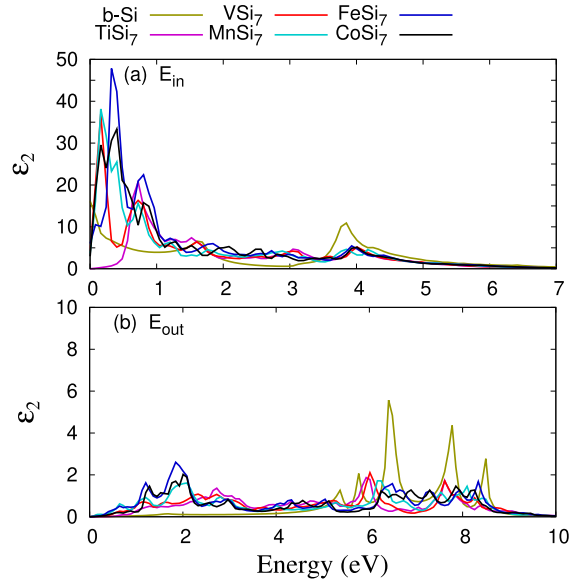}
	\caption{The imaginary dielectric function of pristine b-Si (golden) and TM-silicene including 
		TiSi$_7$ (purple), VSi$_7$ (red), MnSi$_7$ (light blue), FeSi$_7$ (blue), and CoSi$_7$ (black) for in-plane, E$_{\rm in}$, or parallel (a) and out-plane or perpendicular, E$_{\rm out}$, (b) electric fields.}
	\label{fig04}
\end{figure}

The optical response of a 2D structure is also directly related to the electronic band structure \cite{ABDULLAH2019102686}. For instance the imaginary part of the dielectric function denotes the absorbed energy by the structure.
The imaginary dielectric functions, $\varepsilon_2$, in the case of an in-plane or a parallel, 
E$_{\rm in}$, (a) and an out-plane or a perpendicular, E$_{\rm out}$, (b) electric fields are shown in \fig{fig04} for b-Si and TM-silicene. 
It is well known that the two main peaks in the imaginary dielectric function of b-Si are at $1.68$~eV corresponding to the $\pi$ to $\pi^*$ states, and at $3.85$~eV revealing the transition between the
$\sigma$ to the $\sigma^*$ states in the parallel electric field. 
Similar to graphene, inter-band transitions for perpendicular polarized light are observed for pristine b-Si except the transitions here occur below $10$~eV.

It is interesting to see that the optical response for TM-silicene is strong at low energy (in the visible and the UV regions) for both the parallel and the perpendicular polarizations of the electromagnetic fields. This is attributed to the following facts related to the band structure of TM-silicene. 
First, the gap at the $\Gamma$ point in the band structure for all TM-silicene decreases with a increasing atomic radius of the dopant atoms. The higher the atomic radius the smaller gap at $\Gamma$ point is.
Second, parallel bands are formed in all directions between bonding and antibonding orbitals of the TM-silicene. These parallel bands lead to most of the transitions in the visible and the UV regions.
Third, optical transitions occur around van Hove singularities. A van Hove singularity is caused by
a flat band formed along the M to K as is seen in the band structure of MnSi$_7$ and CoSi$_7$. 
The effects of a van Hove singularity on the optical response have been reported for 2D materials~\cite{JOHN2017307}.

\subsection{Spin-dependent calculations}

In this section, we show the results of a spin-dependent model for the pristine b-Si and 
the TM-silicenes. 
First, we preform both ferromagnetic (FM) and antiferromagnetic (AFM) calculations where the strength of the magnetization is assumed to be $0.5$~A/m. The energy difference between the AFM and the FM states is $\Delta E = E_{\rm AFM} - E_{\rm FM}$.  
The $\Delta E$ of TiSi$_7$ is zero indicating a nonmagnetic structure, and for VSi$_7$, 
MnSi$_7$, FeSi$_7$, and CoSi$_7$ the differences are $85.05$, $50.05$, $13.6$, and $-78.9$~meV, respectively. This shows that the VSi$_7$, MnSi$_7$, and FeSi$_7$ favor FM, but 
CoSi$_7$ favors AFM.

Based on $\Delta E$, one can further estimate the Curie temperature, $T^{\rm MFA}_C$, via 
a mean field approximation using 
\begin{equation}
\frac{3}{2} k_B T^{\rm MFA}_C =  \frac{\Delta E}{N_{\rm imp}}, 
\end{equation}
with N$_{\rm imp}$ the number of TM atoms in the structure. The Curie temperature of 
TiSi$_7$, VSi$_7$, MnSi$_7$, FeSi$_7$, and CoSi$_7$ is found to be $0$, $657$, $386$, $105$, and $609$~K. The results for the Curie temperatures in our calculations agree well with a previous study \cite{doi:10.1021/acs.jpclett.7b00115}, and they are candidates for thermospin devices at $100$~K as
our study shows.

\begin{figure}[htb]
	\centering
	\includegraphics[width=0.48\textwidth]{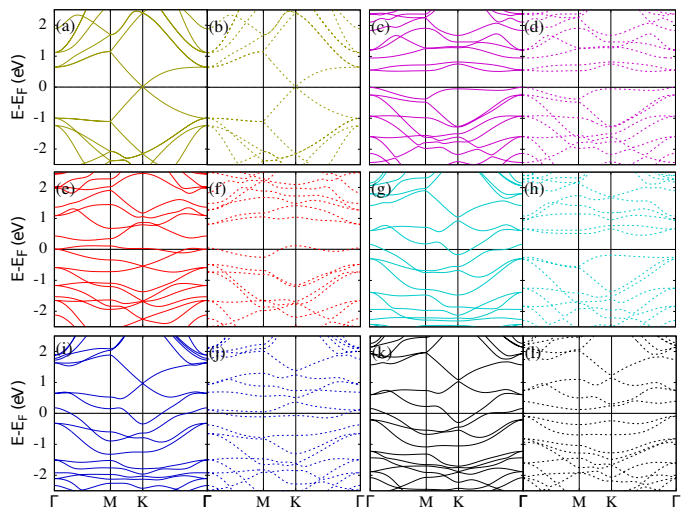}
	\caption{Spin-dependent electronic band structure of b-Si (golden) and TM-silicene including TiSi$_7$ (purple), VSi$_7$ (red), MnSi$_7$ (light blue), FeSi$_7$ (blue), and CoSi$_7$ (black). The sold lines are spin-up and dotted lines are spin-down. The Fermi energy is at 0.}
	\label{fig05}
\end{figure}

\begin{figure}[htb]
	\centering
	\includegraphics[width=0.48\textwidth]{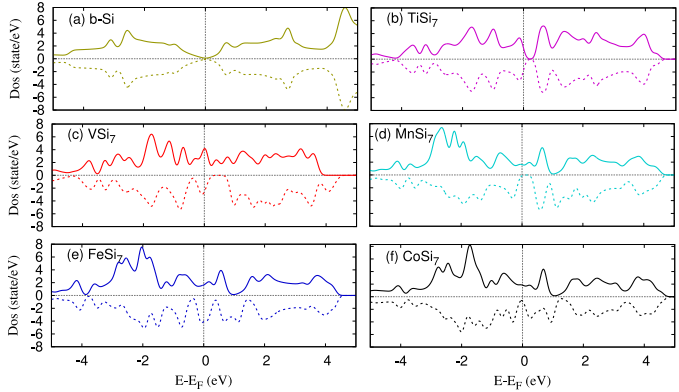}
	\caption{Spin-dependent Dos of b-Si (golden) and TM-silicene including TiSi$_7$ (purple), VSi$_7$ (red), MnSi$_7$ (light blue), FeSi$_7$ (blue), and CoSi$_7$ (black). The sold lines are spin-up and dotted lines are spin-down. The Fermi energy is at 0.}
	\label{fig06}
\end{figure}

It is known that the TM atoms have a strong coupling with silicene giving 
a strong modification of the spin-dependent band structures, density of states, and spin transport properties \cite{doi:10.1021/acs.jpclett.7b00115}. 
Figure \ref{fig05} and \ref{fig06} show the electronic band structure and the Dos of b-Si and TM-silicene for both spin-up (solid lines) and spin-down (dotted lines), respectively.
The band structure and the Dos for b-Si reveals that the spin-up and the spin-down band structures and Dos are symmetric indicating no magnetic properties. In addition, a spin splitting does not occur between the spin-up and the spin-down close to the Fermi energy showing the pristine b-Si is non-magnetic. 

In the TM-silicene, the band structures and the Dos
manifest different trends such as nonmagnetic semiconductors, magnetic metals, and half-metals. 
Namely, TiSi$_7$ (purple) has nonmagnetic or spin unpolarized semiconducting property with bandgap $0.51$~eV for both spin channels. The bandgap here is exactly equal to the bandgap of the structure predicted by the spin-independent model mentioned before.
Spin unpolarized semiconducting is a semiconductors having a bandgap but there is no spin splitting between the spin-up and spin-down, and both spin channels have the same structure. 
These structures are called nonmagnetic semiconductors.
Furthermore, VSi$_7$ (red) and FeSi$_7$ (blue) indicate a magnetic or spin polarized metallic behavior
 with a spin splitting between the spin-up and the spin-down states around the Fermi energy, and the Fermi energy crosses the valence or the conduction bands.
In the two TM-silicene, MnSi$_7$ (light blue) and CoSi$_7$ (black), we notice a spin polarized 
half-metallic feature with the states of one spin direction showing a semiconducting behavior,
while the other displays a metallic behavior. Such property is a basis for spintronic applications. 
The indirect bandgap of the spin-down channel of MnSi$_7$ is $0.64$~eV, and CoSi$_7$ it is $0.21$~eV. 
It is interesting to see that VSi$_7$ has a semiconductor property, when the spin is ignored as 
is shown in \fig{fig02}c, but in the spin-dependent model VSi$_7$ becomes metallic. The same applies
to MnSi$_7$ and CoSi$_7$, which are metallic according to the spin-independent model, but they become half metallic when spin is accounted for. These spin phase transitions have been observed for silicene materials and superlattices \cite{Khoeini2019}.
\begin{figure}[htb]
	\centering
	\includegraphics[width=0.3\textwidth]{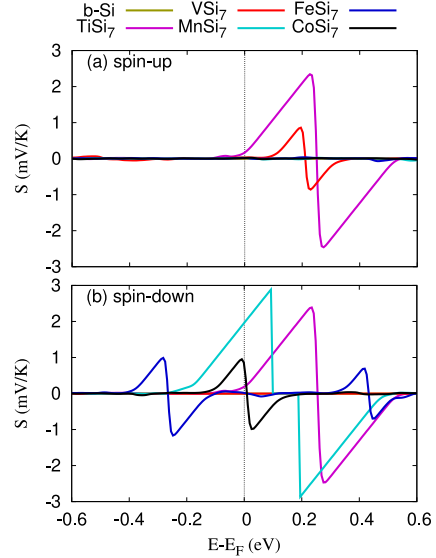}
	\caption{Seebeck coefficient of b-Si and TM-silicene at $T = 100$~K for both spin-up (a) and spin-down (b) channels.}
	\label{fig07}
\end{figure}
The spin-phase transition process leads to an increase in thermal efficiency. 
For instance, the spin-independent Seebeck coefficient of MnSi$_7$ and CoSi$_7$ were small shown in \fig{fig03}, but the spin-polarized Seebeck coefficient of MnSi$_7$ and CoSi$_7$ is enhanced for 
the spin-down channel as is shown in \fig{fig07}. This is attributed to the opening of bandgaps for
the spin-down channel making them half-metals.  It should be mentioned that the Seebeck coefficient of TiSi$_7$ is the same for both spin-up and down because TiSi$_7$ is nonmagnetic material. In contrast, the spin-independent Seebeck coefficient of VSi$_7$ was high but it is suppressed for spin-up and totally vanished for spin-down channel.
This spin-dependent phase transition is an important property, through which a bandgap can be induced 
by means of magnetic dopants. The systems with such a bandgap can be applicable
to practical areas, such as field-effect transistors (FETs) \cite{Tao2015}, single-spin electron sources \cite{Tsai2013}, and nonvolatile magnetic random access memory \cite{Ahn2018}.
This is also important for thermospin filtering in spintronic devices.

\begin{figure}[htb]
	\centering
	\includegraphics[width=0.3\textwidth]{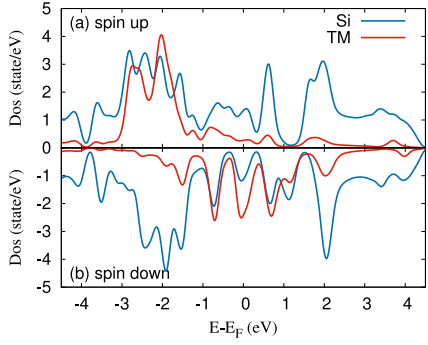}
	\caption{The spin-dependent density of state of the Si and TM atoms in TM-silicene, FeSi$_7$, for spin-up (a) and spin-down (b).}
	\label{fig08}
\end{figure}

The spin-dependent phase transition can mainly be referred to the spin-dependent orbitals 
of the TM atoms, which contribute to the energy level through the hybridization between 
the TM atom and the silicene layer as is presented in \fig{fig08}. We clearly see
that the density of state of the TM atom for both spin-up and spin-down around the Fermi energy 
has high contributions. Especially, the spin-down density of states of TM atoms is much stronger around the Fermi energy compared to the spin-up states.

\begin{figure}[htb]
	\centering
	\includegraphics[width=0.3\textwidth]{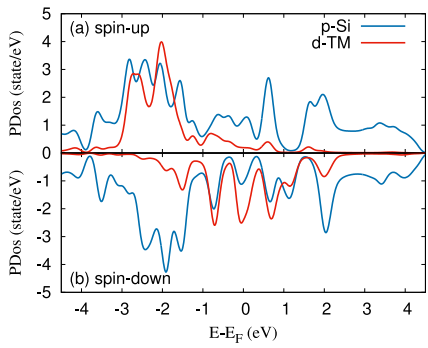}
	\caption{The partial density of state, PDos, of the d-orbital of TM-atom, d-TM, and p-orbital of 
		Si atoms of silicene, p-Si, for both spin-up (a), and spin-down (b).}
	\label{fig09}
\end{figure}

Another magnetic property of the TM-silicene is the magnetic moment. 
The $3d$-orbitals of the TM atoms gives rise to the magnetism of TM-silicene, and its
partial density of state difference for both spin-up and spin-down states shown in \fig{fig09},
which is responsible for the net magnetic moment. 
In particular, one can observe that the PDos (partial Dos) peaks of the d-orbitals of the transition metal, d-TM, can align with the peaks of the p-orbitals of the Si atoms of silicene, p-Si, very well. This implies that there are the strong interaction and large orbital hybridization between the TM 
atoms and the silicene. As a result, high magnetic moments are found in TM-silicene.

\begin{figure}[htb]
	\centering
	\includegraphics[width=0.4\textwidth]{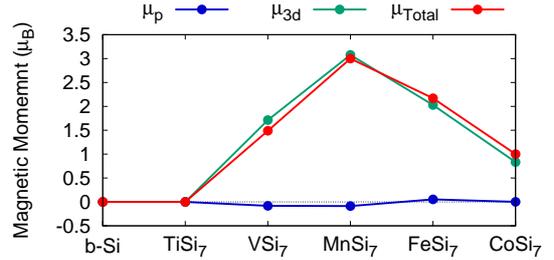}
	\caption{Magnetic moments of b-Si and TM-silicene for the p-orbital contribution of Si atoms ($\mu_{\rm p}$), 3d orbitals of TM atoms ($\mu_{\rm 3d}$), and total magnetization of TM-silicene ($\mu_{\rm Total}$).}
	\label{fig10}
\end{figure}

Figure \ref{fig10} shows the magnetic moments of the b-Si and the TM-silicene systems for the p-orbital contribution of the Si atoms ($\mu_{\rm p}$), 3d orbitals of the TM atoms ($\mu_{\rm 3d}$), and the total magnetization of the TM-silicene ($\mu_{\rm Total}$). One can clearly see that the magnetization is mainly caused by the TM atoms. In addition, we should remember that our results for 
the magnetic moment is underestimated by the GGA-PBE calculations. If GGA$+$U or HSE calculations are used, the obtained magnetic moments will be higher. For instance, the magnetic moment of MnSi$_7$ is 
3$\mu_{\rm B}$ using the GGA-PBE approach, but it is increased to 4$\mu_{\rm B}$ if the GGA$+$U or 
HSE are used. 
This smaller value of magnetic moment of MnSi$_7$ here can be attributed to the fact that, the semilocal GGA functional tends to delocalize the $d$ electrons and increase the $d\text{-}p$ overlapping, consequently leading to the underestimation of the magnetic
moment of the Mn dopant \cite{doi:10.1021/acs.jpclett.7b00115}.

\begin{figure}[htb]
	\centering
	\includegraphics[width=0.5\textwidth]{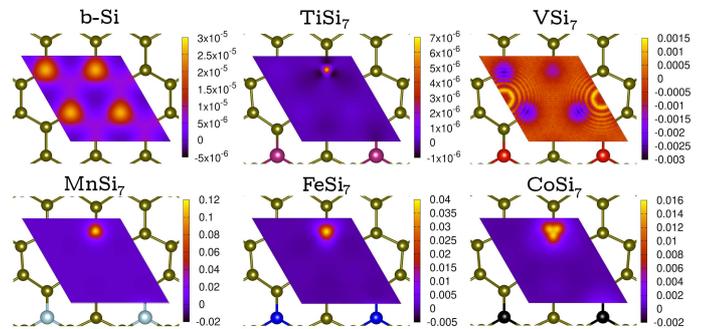}
	\caption{The spin polarized density of b-Si and TM-silicene structures.}
	\label{fig11}
\end{figure}

In order to give a more clear picture of the magnetic distribution, we plot the spin polarized 
density ($\Delta \rho = \rho_{\rm up} - \rho_{\rm down}$) in \fig{fig11} for b-Si and TM-silicene. 
The spin polarized distribution of b-Si and TiSi$_7$ is very small. It is neglectable even on the finer scale used on these two subfigures (see the scale of the z-axis). Remarkably, The spin polarized distribution of other TM-silicenes is almost entirely located
on the dopant atoms, showing highly localized magnetic features.
Especially, the spin polarization of the Mn atom in MnSi$_7$ is highest, revealing the highest 
magnetic moment as was mentioned before.

\section{Summary and Conclusions}\label{Sec:Conclusion}

We have studied TM-silicene with density functional computation models neglecting, or accounting 
for spin-polarization.
In the absence of spin-polarization in TM-silicene, we observe both metallic and semiconductor 
behavior depending on the TM dopant atoms. As a result a thermoelectric property such as the Seebeck coefficient is enhanced, and TM-silicenes show a large optical response at low energy. 
In the presence of spin-polarization in TM-silicene, a spin phase transitions occurs leading to  
half metallic, metallic, and semiconductor properties of the TM-silicene. The spin phase transitions induce spin filtering or the possibility to access a thermospin transport via spin-up and down channels depending on the spin-dependent band structure. In addition, strong orbital interactions between the Silicon and the TM atoms are observed. These interactions control the magnetization of the system, in which a high magnetic moment of TM-silicene is generated.

\section{Acknowledgment}
This work was financially supported by the University of Sulaimani and 
the Research center of Komar University of Science and Technology. 
The computations were performed on resources provided by the Division of Computational 
Nanoscience at the University of Sulaimani.



\begin{thebibliography}{50}
	\providecommand{\url}[1]{#1}
	\csname url@samestyle\endcsname
	\providecommand{\newblock}{\relax}
	\providecommand{\bibinfo}[2]{#2}
	\providecommand{\BIBentrySTDinterwordspacing}{\spaceskip=0pt\relax}
	\providecommand{\BIBentryALTinterwordstretchfactor}{4}
	\providecommand{\BIBentryALTinterwordspacing}{\spaceskip=\fontdimen2\font plus
		\BIBentryALTinterwordstretchfactor\fontdimen3\font minus
		\fontdimen4\font\relax}
	\providecommand{\BIBforeignlanguage}[2]{{%
			\expandafter\ifx\csname l@#1\endcsname\relax
			\typeout{** WARNING: IEEEtran.bst: No hyphenation pattern has been}%
			\typeout{** loaded for the language `#1'. Using the pattern for}%
			\typeout{** the default language instead.}%
			\else
			\language=\csname l@#1\endcsname
			\fi
			#2}}
	\providecommand{\BIBdecl}{\relax}
	\BIBdecl
	
	\bibitem{PhysRevB.50.14916}
	\BIBentryALTinterwordspacing
	K.~Takeda and K.~Shiraishi, ``Theoretical possibility of stage corrugation in
	si and ge analogs of graphite,'' \emph{Phys. Rev. B}, vol.~50, pp.
	14\,916--14\,922, Nov 1994. [Online]. Available:
	\url{https://link.aps.org/doi/10.1103/PhysRevB.50.14916}
	\BIBentrySTDinterwordspacing
	
	\bibitem{Tao2015}
	\BIBentryALTinterwordspacing
	L.~Tao, E.~Cinquanta, D.~Chiappe, C.~Grazianetti, M.~Fanciulli, M.~Dubey,
	A.~Molle, and D.~Akinwande, ``Silicene field-effect transistors operating at
	room temperature,'' \emph{Nature Nanotechnology}, vol.~10, no.~3, pp.
	227--231, Mar 2015. [Online]. Available:
	\url{https://doi.org/10.1038/nnano.2014.325}
	\BIBentrySTDinterwordspacing
	
	\bibitem{Jia2016}
	\BIBentryALTinterwordspacing
	T.-T. Jia, X.-Y. Fan, M.-M. Zheng, and G.~Chen, ``Silicene nanomeshes: bandgap
	opening by bond symmetry breaking and uniaxial strain,'' \emph{Scientific
		Reports}, vol.~6, no.~1, p. 20971, Feb 2016. [Online]. Available:
	\url{https://doi.org/10.1038/srep20971}
	\BIBentrySTDinterwordspacing
	
	\bibitem{ZHAO201624}
	\BIBentryALTinterwordspacing
	J.~Zhao, H.~Liu, Z.~Yu, R.~Quhe, S.~Zhou, Y.~Wang, C.~C. Liu, H.~Zhong, N.~Han,
	J.~Lu, Y.~Yao, and K.~Wu, ``Rise of silicene: A competitive 2d material,''
	\emph{Progress in Materials Science}, vol.~83, pp. 24 -- 151, 2016. [Online].
	Available:
	\url{http://www.sciencedirect.com/science/article/pii/S0079642516300068}
	\BIBentrySTDinterwordspacing
	
	\bibitem{RASHID2019102625}
	\BIBentryALTinterwordspacing
	H.~O. Rashid, N.~R. Abdullah, and V.~Gudmundsson, ``Silicon on a graphene
	nanosheet with triangle- and dot-shape: Electronic structure, specific heat,
	and thermal conductivity from first-principle calculations,'' \emph{Results
		in Physics}, vol.~15, p. 102625, 2019. [Online]. Available:
	\url{http://www.sciencedirect.com/science/article/pii/S2211379719317140}
	\BIBentrySTDinterwordspacing
	
	\bibitem{LeLay2015}
	\BIBentryALTinterwordspacing
	G.~Le~Lay, ``Silicene transistors,'' \emph{Nature Nanotechnology}, vol.~10,
	no.~3, pp. 202--203, Mar 2015. [Online]. Available:
	\url{https://doi.org/10.1038/nnano.2015.10}
	\BIBentrySTDinterwordspacing
	
	\bibitem{Do2019}
	\BIBentryALTinterwordspacing
	T.-N. Do, G.~Gumbs, P.-H. Shih, D.~Huang, C.-W. Chiu, C.-Y. Chen, and M.-F.
	Lin, ``Peculiar optical properties of bilayer silicene under the influence of
	external electric and magnetic fields,'' \emph{Scientific Reports}, vol.~9,
	no.~1, p. 624, Jan 2019. [Online]. Available:
	\url{https://doi.org/10.1038/s41598-018-36547-1}
	\BIBentrySTDinterwordspacing
	
	\bibitem{Fu2015}
	\BIBentryALTinterwordspacing
	H.-H. Fu, D.-D. Wu, Z.-Q. Zhang, and L.~Gu, ``Spin-dependent seebeck effect,
	thermal colossal magnetoresistance and negative differential thermoelectric
	resistance in zigzag silicene nanoribbon heterojunciton,'' \emph{Scientific
		Reports}, vol.~5, no.~1, p. 10547, May 2015. [Online]. Available:
	\url{https://doi.org/10.1038/srep10547}
	\BIBentrySTDinterwordspacing
	
	\bibitem{Tokmachev2018}
	\BIBentryALTinterwordspacing
	A.~M. Tokmachev, D.~V. Averyanov, O.~E. Parfenov, A.~N. Taldenkov, I.~A.
	Karateev, I.~S. Sokolov, O.~A. Kondratev, and V.~G. Storchak, ``Emerging
	two-dimensional ferromagnetism in silicene materials,'' \emph{Nature
		Communications}, vol.~9, no.~1, p. 1672, Apr 2018. [Online]. Available:
	\url{https://doi.org/10.1038/s41467-018-04012-2}
	\BIBentrySTDinterwordspacing
	
	\bibitem{C9RA04602J}
	\BIBentryALTinterwordspacing
	Y.~Sun, A.~Huang, and Z.~Wang, ``Transition metal atom (ti{,} v{,} mn{,} fe{,}
	and co) anchored silicene for hydrogen evolution reaction,'' \emph{RSC Adv.},
	vol.~9, pp. 26\,321--26\,326, 2019. [Online]. Available:
	\url{http://dx.doi.org/10.1039/C9RA04602J}
	\BIBentrySTDinterwordspacing
	
	\bibitem{C9CP04284A}
	\BIBentryALTinterwordspacing
	H.~Dong, Y.~Ji, L.~Ding, and Y.~Li, ``Strategies for computational design and
	discovery of two-dimensional transition-metal-free materials for
	electro-catalysis applications,'' \emph{Phys. Chem. Chem. Phys.}, vol.~21,
	pp. 25\,535--25\,547, 2019. [Online]. Available:
	\url{http://dx.doi.org/10.1039/C9CP04284A}
	\BIBentrySTDinterwordspacing
	
	\bibitem{Zhu_2016}
	\BIBentryALTinterwordspacing
	J.~Zhu and U.~Schwingenschlögl, ``Silicene for na-ion battery applications,''
	\emph{2D Materials}, vol.~3, no.~3, p. 035012, aug 2016. [Online]. Available:
	\url{https://doi.org/10.1088%2F2053-1583%2F3%2F3%2F035012}
		\BIBentrySTDinterwordspacing
		
		\bibitem{zhang2018lithiation}
		W.~Zhang, L.~Sun, J.~M.~V. Nsanzimana, and X.~Wang, ``Lithiation/delithiation
		synthesis of few layer silicene nanosheets for rechargeable li--o2
		batteries,'' \emph{Advanced Materials}, vol.~30, no.~15, p. 1705523, 2018.
		
		\bibitem{Sadeghi2015}
		\BIBentryALTinterwordspacing
		H.~Sadeghi, S.~Sangtarash, and C.~J. Lambert, ``Enhanced thermoelectric
		efficiency of porous silicene nanoribbons,'' \emph{Scientific Reports},
		vol.~5, no.~1, p. 9514, Mar 2015. [Online]. Available:
		\url{https://doi.org/10.1038/srep09514}
		\BIBentrySTDinterwordspacing
		
		\bibitem{PhysRevB.88.165422}
		\BIBentryALTinterwordspacing
		J.~Zhang, B.~Zhao, and Z.~Yang, ``Abundant topological states in silicene with
		transition metal adatoms,'' \emph{Phys. Rev. B}, vol.~88, p. 165422, Oct
		2013. [Online]. Available:
		\url{https://link.aps.org/doi/10.1103/PhysRevB.88.165422}
		\BIBentrySTDinterwordspacing
		
		\bibitem{doi:10.1002/cphc.201402613}
		\BIBentryALTinterwordspacing
		Y.~Lee, K.-H. Yun, S.~B. Cho, and Y.-C. Chung, ``Electronic properties of
		transition-metal-decorated silicene,'' \emph{ChemPhysChem}, vol.~15, no.~18,
		pp. 4095--4099, 2014. [Online]. Available:
		\url{https://chemistry-europe.onlinelibrary.wiley.com/doi/abs/10.1002/cphc.201402613}
		\BIBentrySTDinterwordspacing
		
		\bibitem{Zhang2013}
		\BIBentryALTinterwordspacing
		X.-L. Zhang, L.-F. Liu, and W.-M. Liu, ``Quantum anomalous hall effect and
		tunable topological states in 3d transition metals doped silicene,''
		\emph{Scientific Reports}, vol.~3, no.~1, p. 2908, Oct 2013. [Online].
		Available: \url{https://doi.org/10.1038/srep02908}
		\BIBentrySTDinterwordspacing
		
		\bibitem{PhysRevLett.107.076802}
		\BIBentryALTinterwordspacing
		C.-C. Liu, W.~Feng, and Y.~Yao, ``Quantum spin hall effect in silicene and
		two-dimensional germanium,'' \emph{Phys. Rev. Lett.}, vol. 107, p. 076802,
		Aug 2011. [Online]. Available:
		\url{https://link.aps.org/doi/10.1103/PhysRevLett.107.076802}
		\BIBentrySTDinterwordspacing
		
		\bibitem{Wolf1488}
		\BIBentryALTinterwordspacing
		S.~A. Wolf, D.~D. Awschalom, R.~A. Buhrman, J.~M. Daughton, S.~von Moln{\'a}r,
		M.~L. Roukes, A.~Y. Chtchelkanova, and D.~M. Treger, ``Spintronics: A
		spin-based electronics vision for the future,'' \emph{Science}, vol. 294, no.
		5546, pp. 1488--1495, 2001. [Online]. Available:
		\url{https://science.sciencemag.org/content/294/5546/1488}
		\BIBentrySTDinterwordspacing
		
		\bibitem{doi:10.1063/1.4921699}
		\BIBentryALTinterwordspacing
		X.~Sun, L.~Wang, H.~Lin, T.~Hou, and Y.~Li, ``Induce magnetism into silicene by
		embedding transition-metal atoms,'' \emph{Applied Physics Letters}, vol. 106,
		no.~22, p. 222401, 2015. [Online]. Available:
		\url{https://doi.org/10.1063/1.4921699}
		\BIBentrySTDinterwordspacing
		
		\bibitem{ABDULLAH2020113996}
		\BIBentryALTinterwordspacing
		N.~R. Abdullah, C.-S. Tang, A.~Manolescu, and V.~Gudmundsson, ``The interplay
		of electron–photon and cavity-environment coupling on the electron
		transport through a quantum dot system,'' \emph{Physica E: Low-dimensional
			Systems and Nanostructures}, vol. 119, p. 113996, 2020. [Online]. Available:
		\url{http://www.sciencedirect.com/science/article/pii/S1386947719312445}
		\BIBentrySTDinterwordspacing
		
		\bibitem{JOHN2017307}
		\BIBentryALTinterwordspacing
		R.~John and B.~Merlin, ``Optical properties of graphene, silicene, germanene,
		and stanene from ir to far uv – a first principles study,'' \emph{Journal
			of Physics and Chemistry of Solids}, vol. 110, pp. 307 -- 315, 2017.
		[Online]. Available:
		\url{http://www.sciencedirect.com/science/article/pii/S0022369717300367}
		\BIBentrySTDinterwordspacing
		
		\bibitem{C4RA07976K}
		\BIBentryALTinterwordspacing
		R.~Das, S.~Chowdhury, A.~Majumdar, and D.~Jana, ``Optical properties of p and
		al doped silicene: a first principles study,'' \emph{RSC Adv.}, vol.~5, pp.
		41--50, 2015. [Online]. Available: \url{http://dx.doi.org/10.1039/C4RA07976K}
		\BIBentrySTDinterwordspacing
		
		\bibitem{Hasanzadeh_Tazeh_Gheshlagh_2020}
		\BIBentryALTinterwordspacing
		Z.~H.~T. Gheshlagh, J.~Beheshtian, and S.~Mansouri, ``The electronic and
		optical properties of 3d transition metals doped silicene sheet: A {DFT}
		study,'' \emph{Materials Research Express}, vol.~6, no.~12, p. 126326, jan
		2020. [Online]. Available: \url{https://doi.org/10.1088%2F2053-1591%2Fab6541}
			\BIBentrySTDinterwordspacing
			
			\bibitem{ZHAO20161}
			\BIBentryALTinterwordspacing
			W.~Zhao, Z.~Guo, Y.~Zhang, J.~Ding, and X.~Zheng, ``Enhanced thermoelectric
			performance of defected silicene nanoribbons,'' \emph{Solid State
				Communications}, vol. 227, pp. 1 -- 8, 2016. [Online]. Available:
			\url{http://www.sciencedirect.com/science/article/pii/S0038109815004056}
			\BIBentrySTDinterwordspacing
			
			\bibitem{abdullah2020properties}
			N.~R. Abdullah, H.~O. Rashid, C.-S. Tang, A.~Manolescu, and V.~Gudmundsson,
			``Properties of bsi $ \_6 $ n monolayers derived by first-principle
			computation,'' \emph{arXiv preprint arXiv:2008.03782}, 2020.
			
			\bibitem{C4CP01039F}
			\BIBentryALTinterwordspacing
			K.~Zberecki, R.~Swirkowicz, M.~Wierzbicki, and J.~Barnaś, ``Enhanced
			thermoelectric efficiency in ferromagnetic silicene nanoribbons terminated
			with hydrogen atoms,'' \emph{Phys. Chem. Chem. Phys.}, vol.~16, pp.
			12\,900--12\,908, 2014. [Online]. Available:
			\url{http://dx.doi.org/10.1039/C4CP01039F}
			\BIBentrySTDinterwordspacing
			
			\bibitem{Giannozzi_2009}
			\BIBentryALTinterwordspacing
			P.~Giannozzi, S.~Baroni, N.~Bonini, M.~Calandra, R.~Car, C.~Cavazzoni,
			D.~Ceresoli, G.~L. Chiarotti, M.~Cococcioni, I.~Dabo, A.~D. Corso,
			S.~de~Gironcoli, S.~Fabris, G.~Fratesi, R.~Gebauer, U.~Gerstmann,
			C.~Gougoussis, A.~Kokalj, M.~Lazzeri, L.~Martin-Samos, N.~Marzari, F.~Mauri,
			R.~Mazzarello, S.~Paolini, A.~Pasquarello, L.~Paulatto, C.~Sbraccia,
			S.~Scandolo, G.~Sclauzero, A.~P. Seitsonen, A.~Smogunov, P.~Umari, and R.~M.
			Wentzcovitch, ``{QUANTUM} {ESPRESSO}: a modular and open-source software
			project for quantum simulations of materials,'' \emph{Journal of Physics:
				Condensed Matter}, vol.~21, no.~39, p. 395502, sep 2009. [Online]. Available:
			\url{https://doi.org/10.1088%2F0953-8984%2F21%2F39%2F395502}
				\BIBentrySTDinterwordspacing
				
				\bibitem{giannozzi2017advanced}
				P.~Giannozzi, O.~Andreussi, T.~Brumme, O.~Bunau, M.~B. Nardelli, M.~Calandra,
				R.~Car, C.~Cavazzoni, D.~Ceresoli, M.~Cococcioni \emph{et~al.}, ``Advanced
				capabilities for materials modelling with quantum espresso,'' \emph{Journal
					of Physics: Condensed Matter}, vol.~29, no.~46, p. 465901, 2017.
				
				\bibitem{KOKALJ1999176}
				\BIBentryALTinterwordspacing
				A.~Kokalj, ``Xcrysden---a new program for displaying crystalline structures and
				electron densities,'' \emph{Journal of Molecular Graphics and Modelling},
				vol.~17, no.~3, pp. 176--179, 1999. [Online]. Available:
				\url{http://www.sciencedirect.com/science/article/pii/S1093326399000285}
				\BIBentrySTDinterwordspacing
				
				\bibitem{madsen2006boltztrap-2}
				G.~K. Madsen and D.~J. Singh, ``Boltztrap. a code for calculating
				band-structure dependent quantities,'' \emph{Computer Physics
					Communications}, vol. 175, no.~1, pp. 67--71, 2006.
				
				\bibitem{ABDULLAH2020126578}
				\BIBentryALTinterwordspacing
				N.~R. Abdullah, G.~A. Mohammed, H.~O. Rashid, and V.~Gudmundsson, ``Electronic,
				thermal, and optical properties of graphene like sicx structures: Significant
				effects of si atom configurations,'' \emph{Physics Letters A}, vol. 384,
				no.~24, p. 126578, 2020. [Online]. Available:
				\url{http://www.sciencedirect.com/science/article/pii/S037596012030445X}
				\BIBentrySTDinterwordspacing
				
				\bibitem{ABDULLAH2020126350}
				\BIBentryALTinterwordspacing
				N.~R. Abdullah, H.~O. Rashid, M.~T. Kareem, C.-S. Tang, A.~Manolescu, and
				V.~Gudmundsson, ``Effects of bonded and non-bonded b/n codoping of graphene
				on its stability, interaction energy, electronic structure, and power
				factor,'' \emph{Physics Letters A}, vol. 384, no.~12, p. 126350, 2020.
				[Online]. Available:
				\url{http://www.sciencedirect.com/science/article/pii/S0375960120301602}
				\BIBentrySTDinterwordspacing
				
				\bibitem{ABDULLAH2020126807}
				\BIBentryALTinterwordspacing
				N.~R. Abdullah, H.~O. Rashid, C.-S. Tang, A.~Manolescu, and V.~Gudmundsson,
				``Modeling electronic, mechanical, optical and thermal properties of
				graphene-like bc6n materials: Role of prominent bn-bonds,'' \emph{Physics
					Letters A}, vol. 384, no.~32, p. 126807, 2020. [Online]. Available:
				\url{http://www.sciencedirect.com/science/article/pii/S0375960120306745}
				\BIBentrySTDinterwordspacing
				
				\bibitem{ABDULLAH2020103282}
				\BIBentryALTinterwordspacing
				N.~R. Abdullah, D.~A. Abdalla, T.~Y. Ahmed, S.~W. Abdulqadr, and H.~O. Rashid,
				``Effect of bn dimers on the stability, electronic, and thermal properties of
				monolayer graphene,'' \emph{Results in Physics}, vol.~18, p. 103282, 2020.
				[Online]. Available:
				\url{http://www.sciencedirect.com/science/article/pii/S2211379720317496}
				\BIBentrySTDinterwordspacing
				
				\bibitem{PhysRevB.79.115409}
				\BIBentryALTinterwordspacing
				S.~Leb\`egue and O.~Eriksson, ``Electronic structure of two-dimensional
				crystals from ab initio theory,'' \emph{Phys. Rev. B}, vol.~79, p. 115409,
				Mar 2009. [Online]. Available:
				\url{https://link.aps.org/doi/10.1103/PhysRevB.79.115409}
				\BIBentrySTDinterwordspacing
				
				\bibitem{PhysRevLett.102.236804}
				\BIBentryALTinterwordspacing
				S.~Cahangirov, M.~Topsakal, E.~Akt\"urk, H.~\ifmmode~\mbox{\c{S}}\else
				\c{S}\fi{}ahin, and S.~Ciraci, ``Two- and one-dimensional honeycomb
				structures of silicon and germanium,'' \emph{Phys. Rev. Lett.}, vol. 102, p.
				236804, Jun 2009. [Online]. Available:
				\url{https://link.aps.org/doi/10.1103/PhysRevLett.102.236804}
				\BIBentrySTDinterwordspacing
				
				\bibitem{abdullah2020interlayer}
				N.~R. Abdullah, H.~O. Rashid, A.~Manolescu, and V.~Gudmundsson, ``Interlayer
				interaction controlling the properties of ab-and aa-stacked bilayer
				graphene-like bc $ \_ $\{$14$\}$ $ n and si $ \_ $\{$2$\}$ $ c $ \_
				$\{$14$\}$ $,'' \emph{arXiv preprint arXiv:2008.10888}, 2020.
				
				\bibitem{doi:10.1021/acs.jpclett.7b00115}
				\BIBentryALTinterwordspacing
				S.~Li, Z.~Ao, J.~Zhu, J.~Ren, J.~Yi, G.~Wang, and W.~Liu, ``Strain controlled
				ferromagnetic-antiferromagnetic transformation in mn-doped silicene for
				information transformation devices,'' \emph{The Journal of Physical Chemistry
					Letters}, vol.~8, no.~7, pp. 1484--1488, 2017, pMID: 28301928. [Online].
				Available: \url{https://doi.org/10.1021/acs.jpclett.7b00115}
				\BIBentrySTDinterwordspacing
				
				\bibitem{PhysRevB.87.085423}
				\BIBentryALTinterwordspacing
				H.~Sahin and F.~M. Peeters, ``Adsorption of alkali, alkaline-earth, and 3$d$
				transition metal atoms on silicene,'' \emph{Phys. Rev. B}, vol.~87, p.
				085423, Feb 2013. [Online]. Available:
				\url{https://link.aps.org/doi/10.1103/PhysRevB.87.085423}
				\BIBentrySTDinterwordspacing
				
				\bibitem{PhysRevB.87.241411}
				\BIBentryALTinterwordspacing
				S.~{Yi\ifmmode \breve{g}\else {\u g}\fi{}en}, V.~Tayari, J.~O. Island, J.~M.
				Porter, and A.~R. Champagne, ``Electronic thermal conductivity measurements
				in intrinsic graphene,'' \emph{Phys. Rev. B}, vol.~87, p. 241411, Jun 2013.
				[Online]. Available:
				\url{https://link.aps.org/doi/10.1103/PhysRevB.87.241411}
				\BIBentrySTDinterwordspacing
				
				\bibitem{abdullah2019manifestation}
				N.~R. Abdullah, C.-S. Tang, A.~Manolescu, and V.~Gudmundsson, ``Manifestation
				of the purcell effect in current transport through a dot--cavity--qed
				system,'' \emph{Nanomaterials}, vol.~9, no.~7, p. 1023, 2019.
				
				\bibitem{abdullah2019thermoelectric}
				------, ``Thermoelectric inversion in a resonant quantum dot-cavity system in
				the steady-state regime,'' \emph{Nanomaterials}, vol.~9, no.~5, p. 741, 2019.
				
				\bibitem{doi:10.1063/1.5100985}
				\BIBentryALTinterwordspacing
				E.~H. Hasdeo, L.~P.~A. Krisna, M.~Y. Hanna, B.~E. Gunara, N.~T. Hung, and
				A.~R.~T. Nugraha, ``Optimal band gap for improved thermoelectric performance
				of two-dimensional dirac materials,'' \emph{Journal of Applied Physics}, vol.
				126, no.~3, p. 035109, 2019. [Online]. Available:
				\url{https://doi.org/10.1063/1.5100985}
				\BIBentrySTDinterwordspacing
				
				\bibitem{ABDULLAH2019102686}
				\BIBentryALTinterwordspacing
				N.~R. Abdullah, ``Rabi-resonant and intraband transitions in a multilevel
				quantum dot system controlled by the cavity-photon reservoir and the
				electron-photon coupling,'' \emph{Results in Physics}, vol.~15, p. 102686,
				2019. [Online]. Available:
				\url{http://www.sciencedirect.com/science/article/pii/S221137971932251X}
				\BIBentrySTDinterwordspacing
				
				\bibitem{Khoeini2019}
				\BIBentryALTinterwordspacing
				F.~Khoeini and Z.~Jafarkhani, ``Tunable spin transport and quantum phase
				transitions in silicene materials and superlattices,'' \emph{Journal of
					Materials Science}, vol.~54, no.~23, pp. 14\,483--14\,494, Dec 2019.
				[Online]. Available: \url{https://doi.org/10.1007/s10853-019-03928-4}
				\BIBentrySTDinterwordspacing
				
				\bibitem{Tsai2013}
				\BIBentryALTinterwordspacing
				W.-F. Tsai, C.-Y. Huang, T.-R. Chang, H.~Lin, H.-T. Jeng, and A.~Bansil,
				``Gated silicene as a tunable source of nearly 100{\%} spin-polarized
				electrons,'' \emph{Nature Communications}, vol.~4, no.~1, p. 1500, Feb 2013.
				[Online]. Available: \url{https://doi.org/10.1038/ncomms2525}
				\BIBentrySTDinterwordspacing
				
				\bibitem{Ahn2018}
				\BIBentryALTinterwordspacing
				E.~C. Ahn, H.-S.~P. Wong, and E.~Pop, ``Carbon nanomaterials for non-volatile
				memories,'' \emph{Nature Reviews Materials}, vol.~3, no.~3, p. 18009, Mar
				2018. [Online]. Available: \url{https://doi.org/10.1038/natrevmats.2018.9}
				\BIBentrySTDinterwordspacing
				
			\end{thebibliography}


\end{document}